\documentclass[12pt]{article}
\usepackage{amsmath}
\usepackage{graphicx,psfrag,epsf}
\usepackage{enumerate}
\usepackage{natbib}
\usepackage{url} 
\usepackage{appendix}

\newcommand{\blind}{0}

\begin{document}

\def\spacingset#1{\renewcommand{\baselinestretch}%
{#1}\small\normalsize} \spacingset{1}


\if0\blind
{
  \title{\bf Linear Regression under Special Relativity}
  \author{Si Hyung Joo\\
    (innovation@jnu.ac.kr) \\
    Department of Industrial Engineering, \\ 
    Chonnam National University
    }
  \maketitle
} \fi

\if1\blind
{
  \bigskip
  \bigskip
  \bigskip
  \begin{center}
    {\LARGE\bf Title}
\end{center}
  \medskip
} \fi

\bigskip
\begin{abstract}
This study investigated the problem posed by using ordinary least squares (OLS) to estimate parameters of simple linear regression under a specific context of special relativity, where an independent variable is restricted to an open interval, $(-c, c)$. It is found that the OLS estimate for the slope coefficient is not invariant under Lorentz velocity transformation. Accordingly, an alternative estimator for the parameters of linear regression under special relativity is proposed. This estimator can be considered a generalization of the OLS estimator under special relativity; when $c$ approaches to infinity, the proposed estimator and its variance converges to the OLS estimator and its variance, respectively. The variance of the proposed estimator is larger than that of the OLS estimator, which implies that hypothesis testing using the OLS estimator and its variance may result in a liberal test under special relativity.
\end{abstract}

\noindent%
\vfill

\newpage
\spacingset{2}
\section{Introduction}
\label{sec:intro}

Linear regression is one of the most frequently employed models in empirical analysis.
\begin{equation*}
Y_{i}=\beta_{0}+\beta_{1} \cdot X_{1i}+\beta_{2} \cdot X_{2i} + \cdots + \beta_{p} \cdot X_{pi} + \epsilon_{i}
\end{equation*}

The $\beta_{p}$ is a measure of association between the independent variable $X_{p}$ and the dependent variable $Y$, and $\beta_{0}$ is the expected value of $Y$ when all $X_{p}$s are equal to zero. The $\epsilon_{i}$ is the error term.

The linear regression model is based on the following assumptions. 

(1) The independent variables are measured without error.

(2) The errors are independent from the independent variables.

(3) The errors are independently and identically normally distributed.

The unknown parameters in a linear regression model are often estimated using the ordinary least squares (OLS) method because the OLS estimator has desirable properties as an estimator of parameters, such as unbiasedness, consistency, and efficiency \citep{r1}. 

The linear regression model and OLS estimator provide accurate inferences and estimates only if the assumptions above hold true. The assumption of an error term that is normally distributed conditional on the independent variables implies that the dependent variable can be any real number. This assumption is violated if a dependent variable has a limited range, for example, a discontinuous or bounded dependent variable. Because a linear regression model with a limited dependent variable may lead to serious errors of inference, alternative nonlinear models and procedures have been developed and employed, such as a Tobit model for censored dependent variables and a Poisson regression model for count (non-negative integer) dependent variables \citep{r2}.

In contrast, researchers rarely pay attention to whether independent variables with a limited range exist in the model---as long as they are exogenous and measured without error---because no assumption of the linear regression model is violated. Although they have limited independent variables, conventional (e.g., OLS) estimators are commonly used to estimate the unknown parameters of the linear regression model.

Are there no problems posed by using the OLS estimator when independent variables are restricted as long as they are exogenous and measured without error? If there are, what would be the proper estimator when independent variables are restricted? 

To investigate the problem posed by using the OLS estimator when an independent variable is restricted, this study investigates a simple linear regression model with an exogenous error-free independent variable intrinsically restricted to an open interval. The linear regression model emerges when one tries to estimate the scale (slope coefficient) and the accuracy (intercept coefficient) of a velocity meter under special relativity (see Section 2 for details). In this model, the dependent variable is the velocity of an object measured by an observer with a velocity meter that has a normal error, and the independent variable is the true velocity of the object relative to the observer. In the real world, where the special theory of relativity applies, the true velocity of a (massive) object from an observer (independent variable) is restricted to an open interval, $(-c,c)$, where c is the speed of light \citep{r3}.

The OLS estimate for the slope coefficient is found to depend on the velocity of an observer under special relativity. To address the problem, a new estimator for the slope coefficient, which is independent from the velocity of an observer, is proposed. 

The proposed estimator is found to be unbiased and converges to the OLS estimator when $c$ approaches to infinity. Its variance is larger than the OLS estimator, which reflects the fact that there is larger uncertainty if an independent variable is restricted. Its variance also converges to that of the OLS estimator when $c$ approaches to infinity.

The rest of this paper is organized as follows. Section 2 illustrates the linear regression model under special relativity. Section 3 investigates the inadequacy of using the OLS estimator for the linear regression model under special relativity. Section 4 provides the rationale for an alternative estimator. Section 5 proposes an alternative estimator for the linear regression model under special relativity. Section 6 examines the properties of the alternative estimator. Finally, Section 7 summarizes and provides concluding remarks.

\section{Linear regression model under special relativity}
\label{sec:LRMSR}

In this section, we describe a special relativistic situation in which a simple linear regression model with an independent variable that is intrinsically bounded $(-c, c)$ emerges.

Suppose an engineer developed a velocity meter. In terms of precision, the velocity meter has a random error that follows a normal distribution with zero mean and unknown variance $\sigma^{2}$. The random error is independent not only from the true velocity of the object being measured but also from the velocity of the observer holding the velocity meter. The engineer is not sure through which unit---meter/second, mile/hour, or others---their velocity meter measures the velocity of an object. If the true velocity of an object is measured by meter/second, the unit of the newly developed velocity meter can be represented by $\beta_{1}$ meter/second. In the worst case, their velocity meter does not reflect the true velocity of objects at all, that is, $\beta_{1}=0$. In addition, the engineer is not sure if their velocity meter is zero adjusted ($\beta_{0}=0$), that is, if the velocity meter shows zero velocity when measuring the velocity of a stationary object from the observer. In other words, the velocity meter has a systematic bias $\beta_{0}$ in terms of accuracy. 

The engineer took their velocity meter to their fellow researcher who has a velocity meter that exactly measures the velocity of an object in meter/second. The engineer asked the researcher if their velocity meter can measure the velocity of an object---that is, $\beta_{1} \neq 0$---and, if so, which scale the velocity meter is using ($\beta_{1}=0$) and how much their own velocity meter needs to be adjusted ($\beta_{0}=0$) to ensure zero velocity for a stationary object. In the real (relativistic) world, the true velocity of a (massive) object is restricted to an open interval, $(-c,c)$, where $c$ is the speed of light in meter/second \citep{r3}. The researcher's velocity meter therefore always shows a value within a range $(-c,c)$.

Therefore, the population regression equation can be represented as
\begin{equation} 
Y_{i}=\beta_{0}+\beta_{1} \cdot X_{i}+ \epsilon_{i} \label{eq01}
\end{equation}
where:
\begin{verse}
$Y_{i}$: the velocity of object $i$ measured by the newly developed velocity meter \\
$X_{i}$: the true velocity of object $i$ in terms of meter/second, $X_{i} \in (-c, c)$ \\
$\beta_{0}$: systematic bias of the velocity meter for a stationary object \\
$\beta_{1}$: the scale of the velocity meter in terms of meter/second \\
$\epsilon_{i}$: random error of the velocity meter, which follows $N(0,\sigma^{2})$ 
\end{verse}

\section{Investigating the inadequacy of OLS estimator under special relativity}

In this section, the inadequacy of using the OLS estimator under special relativity is investigated.

\subsection{Newtonian universe}
Let us suppose the engineer and researcher were living in the Newtonian universe. Because the true velocity of an (massive) object can range from $-\infty$ to $\infty$ in the Newtonian universe \citep{r3}, the regression model becomes a simple linear regression model with an unrestricted independent variable. 

The researcher may conduct an experiment measuring the velocity of $N$ objects moving along a straight line with both a newly developed velocity meter and the exact velocity meter. With the relevant data, they can estimate unknown parameters using the OLS estimator. 

The OLS sample regression equation corresponding to equation \eqref{eq01} can be written as
\begin{equation} 
Y_{i}=\hat{\beta}_{0, OLS} + \hat{\beta}_{1, OLS} \cdot X_{i} + \hat{\epsilon}_{i}
\end{equation}
where $\hat{\beta}_{0, OLS}$ and $\hat{\beta}_{1, OLS}$ are the OLS estimator of $\beta_{0}$ and $\beta_{1}$, respectively, and $\hat{\epsilon}_{i}$ is the OLS residual for sample $i$.

The first-order conditions for the OLS estimators $\hat{\beta}_{0, OLS}$ and $\hat{\beta}_{1, OLS}$ are $\sum_{i=1}^{N} \hat{\epsilon}_{i} = 0$
and $\sum_{i=1}^{N} \hat{\epsilon}_{i} \cdot X_{i} = 0$, respectively. 

The OLS estimator can be regarded as a method of moments estimator based on the population moment condition $E[\epsilon]=0$ and $E[\epsilon \cdot X]=0$ \citep{r4}.

The estimates $\hat{\beta}_{0, OLS}$ and $\hat{\beta}_{1, OLS}$ are bivariate normally distributed, and their means, variances, and covariance are as follows.

\begin{eqnarray} 
\hat{\beta}_{1, OLS} 
&=& \frac{\sum_{i=1}^{N} (Y_{i}-\bar{Y}) \cdot ( X_{i}-\bar{X} )}
{\sum_{i=1}^{N} (X_{i}-\bar{X} )^{2}}  \\
\hat{\beta}_{0, OLS} &=& \bar{Y}-\hat{\beta}_{1, OLS} \cdot \bar{X} \\
Var(\hat{\beta}_{1, OLS} ) &=&
\frac{1}{\sum_{i=1}^{N} (X_{i}-\bar{X} )^{2}}\sigma^{2} \\
Var(\hat{\beta}_{0, OLS} ) &=&
\frac{\sum_{i=1}^{N} X_{i}^{2}}{\sum_{i=1}^{N} N(X_{i}-\bar{X} )^{2}}\sigma^{2} \\
Cov(\hat{\beta}_{0, OLS} , \hat{\beta}_{1, OLS} ) &=&
-\frac{\bar{X}}{\sum_{i=1}^{N} (X_{i}-\bar{X} )^{2}}\sigma^{2}
\end{eqnarray}

where $\bar{X}=\frac{1}{N}\sum_{i=1}^{N} X_{i}$ and $\bar{Y}=\frac{1}{N}\sum_{i=1}^{N} Y_{i}$.

$\hat{\beta}_{1, OLS}$ is the estimate of the scale of the velocity meter and $\hat{\beta}_{0, OLS}$ is the estimate of the systematic bias. Their (simultaneous) confidence intervals can be determined by their variances and covariance.

If the researcher were measuring the velocity of objects with the exact velocity meter while moving in a relatively positive direction with a constant velocity $v_{*}$ than before, they would obtain $X'_{i}=X_{i}-v_{*}$ according to the Galilean velocity transformation \citep{r5}. The relationships between $x_{i}$ and $x'_{i}$ can be denoted as follows.

\begin{eqnarray} 
X_{i}&=&(X_{i}-\bar{X})+\bar{X}=x_{i}+\bar{X} \\
X'_{i}&=&X_{i}-v_{*}=x_{i}+\bar{X}-v_{*}=x_{i}+\bar{X'}
\end{eqnarray}
where $x_{i}=X_{i}-\bar{X}$ and $\bar{X'}=\frac{1}{N}\sum_{i=1}^{N} X'_{i}$.

$X_{i}$ and $X'_{i}$ have the same demeaned velocity $x_{i}$. In other words, the demeaned velocity is invariant under Galilean velocity transformation. In addition, $\sum_{i=1}^{N} x_{i}=0$ and $X_{i}$ and $X'_{i}$ and $x_{i}$ have the same variance.

If the researcher were measuring the velocity of objects with the newly developed velocity meter while moving in a positive direction with a constant velocity $v_{*}$ than before, they would obtain $Y'_{i}=Y_{i}-\beta_{1} \cdot (X_{i}-X'_{i}) = Y_{i}-\beta_{1} \cdot v_{*}$. 

Because the error terms need to be independent from the velocity of the researcher as well as the true velocity of objects, the independence of error terms from the true velocity needs to be specified by the demeaned velocity $x_{i}$, which is invariant from the velocity of the researcher. Hence, the independence of error terms from the true velocity of objects and that of the researcher is specified as $E[\epsilon \cdot x]=0$. Because $E[\epsilon \cdot x]=0$ is equivalent to $E[\epsilon \cdot X]=0$ and $E[\epsilon \cdot X']=0$, the error term, which is independent from $x$, is also independent from $X$ and $X'$.

The OLS estimates $\hat{\beta'}_{1, OLS}$ and its variance $Var(\hat{\beta'}_{1, OLS})$ based on $X'_{i}$ and $Y'_{i}$ are the same as $\hat{\beta}_{1, OLS}$ and $Var(\hat{\beta}_{1, OLS})$, respectively. 

It shows that the OLS estimates for $\beta_{1}$ and its variance remain invariant regardless of the velocity of the researcher in the Newtonian universe. In other words, OLS estimates for $\beta_{1}$ and its variance are invariant under the Galilean velocity transformation.

\subsection{Real relativistic world}
The real world is not like the Newtonian universe. In the real (relativistic) world, the true velocity of an (massive) object is restricted to an open interval, $(-c,c)$, where $c$ is the speed of light \citep{r3}. Moreover, if the researcher in the real (relativistic) world were to measure the velocity of objects while moving in a positive direction with a constant velocity $v_{*}$ than before, they would obtain the following velocities according to the Lorentz velocity transformation \citep{r3}.

\begin{eqnarray} 
X''_{i} &=&
\frac{X_{i}-v_{*}}{1-\frac{v_{*} \cdot X_{i}}{c^{2}}} \\
Y''_{i} &=& y_{i} - \beta_{1} \cdot (X_{i}-X''_{i})
\end{eqnarray}

Unlike the Newtonian universe case, the OLS estimate $\hat{\beta''}_{1, OLS}$ and its variance $Var(\hat{\beta''}_{1, OLS})$ based on $X''_{i}$ and $Y''_{i}$ are different from $\hat{\beta}_{1, OLS}$ and $Var(\hat{\beta}_{1, OLS})$. It shows that the OLS estimate for $\beta_{1}$ and its variance depend on the velocity of the researcher in the real (relativistic) world. 

This problem arises because the demeaned velocity is not invariant under the Lorentz velocity transformation. Because $X_{i}$ and $X''_{i}$ have different demeaned velocity, $x_{i} \neq x''_{i}$, where $x''_{i}=X''_{i}-\bar{X''}$ and $\bar{X''}=\frac{1}{N}\sum_{i=1}^{N} X''_{i}$, $E[\epsilon \cdot X]=0$ is not equivalent to $E[\epsilon \cdot X'']=0$.

This result shows that, to obtain an estimate for $\beta_{1}$ and its variance, which is independent from the velocity of the researcher in the real (relativistic) world, the independence of the error term from the velocity of objects and from that of the researcher needs to be specified by a quantity that is invariant under the Lorentz velocity transformation.

\section{Independence of error term in the relativistic universe}
In this section, we search for invariant quantities \citep{r3} under the Lorentz velocity transformation and suggest specifications for the independence of the error term from the velocity of objects and that of the researcher; the goal is to properly estimate the parameters of the regression model. Please refer to \cite{r4} for details of the concepts under special relativity and the Lorentz invariant quantities.

In physics, the rapidity $\theta$ of a velocity $X$ is defined as follows. 

\begin{equation} 
\theta = tanh^{-1}(\frac{X}{c})
\end{equation} 
The relativistic momentum and energy of an object with rapidity $\theta$ and rest mass $m$ are defined as follows.
\begin{eqnarray} 
Momentum &=& m \cdot c \cdot sinh(\theta) \\
Energy &=& m \cdot c^{2} \cdot cosh(\theta)
\end{eqnarray} 
Let $\theta_{i}$, $\theta''_{i}$ and $\theta_{*}$ be the rapidity of $X_{i}$, $X''_{i}$ and $v_{*}$, respectively. 
\begin{eqnarray} 
\theta_{i} &=& tanh^{-1}(\frac{X_{i}}{c}) \\
\theta''_{i} &=& tanh^{-1}(\frac{X''_{i}}{c}) \\
\theta_{*} &=& tanh^{-1}(\frac{v_{*}}{c})
\end{eqnarray} 
Then, the following relationship among $\theta_{i}$, $\theta''_{i}$ and $\theta_{*}$ holds true.
\begin{equation}
\theta''_{i} = \theta_{i} - \theta_{*}
\end{equation}
Hence, $X_{i}$ and $X''_{i}$ can be represented as follows.
\begin{eqnarray}
X_{i} &=& c \cdot tanh(\theta_{i}) \\
X''_{i} &=& c \cdot tanh(\theta''_{i}) = c \cdot tanh(\theta_{i}-\theta_{*})
\end{eqnarray}
Let $\theta_{0}$ and $\theta''_{0}$ be defined as follows.
\begin{eqnarray}
tanh(\theta_{0}) &=& \frac{\sum_{i=1}^{N} sinh(\theta_{i})}{\sum_{i=1}^{N} cosh(\theta_{i})} \\
tanh(\theta''_{0}) &=& \frac{\sum_{i=1}^{N} sinh(\theta''_{i})}{\sum_{i=1}^{N} cosh(\theta''_{i})} \\
\end{eqnarray}
Then, the following relationship among $\theta_{0}$, $\theta''_{0}$ and $\theta_{*}$ holds true.
\begin{equation}
\theta_{*} = \theta_{0} - \theta''_{0}
\end{equation}
Let $\phi_{i}=\theta_{i}-\theta_{0}$; then, $\theta_{i}$ and $\theta''_{i}$ can be represented as follows.
\begin{eqnarray}
\theta_{i} &=& (\theta_{i}-\theta_{0})+\theta_{0}=\phi_{i}+\theta_{0} \\
\theta''_{i} &=& \theta_{i} - \theta_{*} = \theta_{i} - (\theta_{0} - \theta''_{0}) = (\theta_{i} - \theta_{0}) + \theta''_{0} = \phi_{i} + \theta''_{0}
\end{eqnarray}
This result shows that $\phi_{i}$ remains invariant regardless of the velocity of the researcher. 
Because $\phi_{i}$ remains invariant, any function of $\phi_{i}$ --- especially the relativistic momentum $m \cdot c \cdot sinh(\phi_{i})$ and the relativistic energy $m \cdot c^{2} \cdot cosh(\phi_{i})$ --- remain invariant, regardless of the velocity of the researcher.
In addition, the following relationships hold true.
\begin{eqnarray}
\sum_{i=1}^{N} sinh(\phi_{i})
= \sum_{i=1}^{N} sinh(\theta_{i}-\theta_{0})
= \sum_{i=1}^{N} sinh(\theta''_{i}-\theta''_{0})=0 \label{eq04} \\
\sum_{i=1}^{N} cosh(\phi_{i})
= \sum_{i=1}^{N} cosh(\theta_{i}-\theta_{0})
= \sum_{i=1}^{N} cosh(\theta''_{i}-\theta''_{0})
\end{eqnarray}

\eqref{eq04} shows that if the researcher were measuring the velocity (rapidity) of objects while moving in a positive direction with a constant rapidity of $\theta_{0}$ than before, then the sum of the relativistic momentum of objects equals zero if we assume that all the objects have the same rest mass. Under this assumption, the rapidity $\theta_{0}$ is associated with the relativistic center of momentum. Therefore, $\phi_{i}$ can be considered the rapidity of object $i$ measured from the relativistic center of momentum when we assume that all the objects have the same rest mass.

Because the error terms need to be independent not only from the true velocity of an object but also from that of the researcher, the independence of the error term needs to be specified by a quantity that is invariant from the velocity of the researcher. Hence, the independence of the error term can be specified as $E[\epsilon \cdot f(\phi)]=0$. 

The parameters can be estimated using the following sample moment conditions corresponding to the population moment conditions $E[\epsilon]=0$ and $E[\epsilon \cdot f(\phi)]=0$. 

\begin{equation}
\begin{split}
\bar{Y} - \hat{\beta}_{0} - \hat{\beta}_{1} \cdot \bar{X} = 0 \\
\frac {1}{N} \{ \sum_{i=1}^{N} Y_{i} \cdot f(\phi_{i}) - \hat{\beta}_{0} \sum_{i=1}^{N} f(\phi_{i}) - \hat{\beta}_{1} \sum_{i=1}^{N} X_{i} \cdot f(\phi_{i}) \} = 0
\end{split}
\end{equation}

When $\beta_{1} = 0$, the $\beta_{0}$ is not uniquely identified if $\sum_{i=1}^{N} f(\phi_{i}) \neq 0$. $\sum_{i=1}^{N} f(\phi_{i})$ needs to be equal to zero. Therefore, $E[\epsilon \cdot sinh(\phi)]=0$ is selected as the population moment condition.

\section{Special relativistic linear regression estimator}
The population regression equation is the same as equation \eqref{eq01}.
The population moment conditions are 
\begin{eqnarray}
E[\epsilon]=0 \\
E[\epsilon \cdot sinh(\phi)]=0
\end{eqnarray}

The sample regression equation is
\begin{equation} 
Y_{i}=\hat{\beta}_{0}+\hat{\beta}_{1} \cdot X_{i}+ \hat{\epsilon}_{i} 
\end{equation}
Meanwhile, the sample moment conditions are
\begin{eqnarray}
\frac {1}{N}\sum_{i=1}^{N} \hat{\epsilon}_{i}=0 \label{eq02} \\
\frac {1}{N}\sum_{i=1}^{N} \hat{\epsilon}_{i} \cdot sinh(\phi_{i})=0 \label{eq03}
\end{eqnarray}

From equation \eqref{eq02},
\begin{equation}
\begin{split}
\frac {1}{N}\sum_{i=1}^{N} \hat{\epsilon}_{i} &=
\frac {1}{N}\sum_{i=1}^{N} (Y_{i}-\hat{\beta}_{0}-\hat{\beta}_{1} \cdot X_{i}) \\ &=
\frac {1}{N}\sum_{i=1}^{N} Y_{i} -\hat{\beta}_{0} - \frac {\hat{\beta}_{1}}{N}\sum_{i=1}^{N} X_{i} \\ &= 
\bar{Y} - \hat{\beta}_{0} - \hat{\beta}_{1} \cdot \bar{X} = 0
\end{split}
\end{equation}

From equation \eqref{eq03}, 
\begin{align}
\frac {1}{N}\sum_{i=1}^{N} \hat{\epsilon}_{i} \cdot sinh(\phi_{i})
 &= \frac {1}{N}\sum_{i=1}^{N} (Y_{i}-\hat{\beta}_{0}-\hat{\beta}_{1} \cdot X_{i}) \cdot sinh(\phi_{i}) \\ 
&= \frac {1}{N} \{ \sum_{i=1}^{N} Y_{i} \cdot sinh(\phi_{i}) - \hat{\beta}_{0} \sum_{i=1}^{N} sinh(\phi_{i}) - \hat{\beta}_{1} \sum_{i=1}^{N} X_{i} \cdot sinh(\phi_{i}) \} \nonumber \\ 
&= \frac {1}{N} \{ \sum_{i=1}^{N} Y_{i} \cdot sinh(\phi_{i}) - \hat{\beta}_{1} \sum_{i=1}^{N} X_{i} \cdot sinh(\phi_{i}) \} = 0 & \text{by \eqref{eq04}} \nonumber
\end{align}

Note that $\sum_{i=1}^{N} X_{i} \cdot sinh(\phi_{i}) > 0$ (see Appendix for proof).

Therefore,
\begin{eqnarray}
\hat{\beta}_{1} &=& \frac{\sum_{i=1}^{N} Y_{i} \cdot sinh(\phi_{i})}{\sum_{i=1}^{N} X_{i} \cdot sinh(\phi_{i})} \label{eq05}  \\
\hat{\beta}_{0} &=& \bar{Y} - \hat{\beta}_{1} \cdot \bar{X} 
= \bar{Y} - \frac{\sum_{i=1}^{N} Y_{i} \cdot sinh(\phi_{i})}{\sum_{i=1}^{N} X_{i} \cdot sinh(\phi_{i})} \cdot \bar{X} \label{eq07}
\end{eqnarray}

\section{Properties of Special relativistic linear regression estimator}
In this section, the properties of the proposed estimator are examined.
\subsection{Linearity of $\hat{\beta}_{1}$ and $\hat{\beta}_{0}$}
The estimator $\hat{\beta}_{1}$ and $\hat{\beta}_{0}$ can be written as a linear combination of the sample values of $Y$, the $Y_{i}$ $(i=1,\cdots ,N)$.
Note equation \eqref{eq05} and \eqref{eq07}, $\hat{\beta}_{1} = \sum_{i=1}^{N} k_{i} \cdot Y_{i}$ and $\hat{\beta}_{0} = \sum_{i=1}^{N} h_{i} \cdot Y_{i}$, where $k_{i}=\frac{sinh(\phi_{i})}{\sum_{i=1}^{N} X_{i} \cdot sinh(\phi_{i})}$ and $h_{i}=\frac{sinh(\phi_{i})\cdot \bar{X}}{\sum_{i=1}^{N} X_{i} \cdot sinh(\phi_{i})}$.

Because $\hat{\beta}_{1}$ and $\hat{\beta}_{0}$ are linear combination of normally distributed random variables $Y_{i}$, $\hat{\beta}_{1}$ and $\hat{\beta}_{0}$ are normally distributed.  

\subsection{Unbiasedness of $\hat{\beta}_{1}$ and $\hat{\beta}_{0}$}
Note that $ \sum_{i=1}^{N} k_{i} = \frac{\sum_{i=1}^{N} sinh(\phi_{i})}{\sum_{i=1}^{N} X_{i} \cdot sinh(\phi_{i})}=0$ and $ \sum_{i=1}^{N} k_{i} \cdot X_{i} =\frac{\sum_{i=1}^{N} X_{i} \cdot sinh(\phi_{i})}{\sum_{i=1}^{N} X_{i} \cdot sinh(\phi_{i})}=1$.

\begin{equation}
\begin{split}
\hat{\beta}_{1} &= \sum_{i=1}^{N} k_{i} \cdot Y_{i}
= \sum_{i=1}^{N} k_{i} (\beta_{0}+\beta_{1} \cdot X_{i}+ \epsilon_{i}) \\ &=
\beta_{0} \sum_{i=1}^{N} k_{i} + \beta_{1} \sum_{i=1}^{N} k_{i} \cdot X_{i} + \sum_{i=1}^{N} k_{i} \cdot \epsilon_{i} \\ &=
\beta_{1} + \sum_{i=1}^{N} k_{i} \cdot \epsilon_{i} \label{eq06}
\end{split}
\end{equation}

\begin{align}
E[\hat{\beta}_{1}] &= E[\beta_{1} + \sum_{i=1}^{N} k_{i} \cdot \epsilon_{i}]
= E[\beta_{1}] + E[\sum_{i=1}^{N} k_{i} \cdot \epsilon_{i}] \label{eq08} \\ &=
\beta_{1} + \sum_{i=1}^{N} k_{i} \cdot E[\epsilon_{i}|X_{i}] && \text{since $\beta_{1}$ is a constant and the $k_{i}$ are random} \nonumber \\ &=
\beta_{1} + \sum_{i=1}^{N} k_{i} \cdot 0 && \text{since $E[\epsilon_{i}|X_{i}]=0$ by assumption} \nonumber \\ &= \beta_{1} \nonumber
\end{align}

Therefore, $\hat{\beta}_{1}$ is an unbiased estimator of $\beta_{1}$.

\begin{equation}
\hat{\beta}_{0} = \bar{Y} - \hat{\beta}_{1} \cdot \bar{X} 
= (\beta_{0} + \beta_{1} \cdot \bar{X} + \bar{\epsilon} ) - \hat{\beta}_{1} \cdot \bar{X}
= \beta_{0} + (\beta_{1}-\hat{\beta}_{1}) \cdot \bar{X} + \bar{\epsilon}
\end{equation}

\begin{align}
E[\hat{\beta}_{0}] &= E[\beta_{0} + (\beta_{1}-\hat{\beta}_{1}) \cdot \bar{X} + \bar{\epsilon}] \\
&= E[\beta_{0}] + E[(\beta_{1}-\hat{\beta}_{1}) \cdot \bar{X}] + E[\bar{\epsilon}] \nonumber \\
&= \beta_{0} + \bar{X} \cdot E[(\beta_{1}-\hat{\beta}_{1})] +  E[\bar{\epsilon}] && \text{since $\beta_{0}$ is a constant} \nonumber \\
&= \beta_{0} + \bar{X} \cdot E[(\beta_{1}-\hat{\beta}_{1})] && \text{since $E[\bar{\epsilon}]=0$ by assumption} \nonumber \\
&=\beta_{0} + \bar{X} ( E[(\beta_{1}]-E[\hat{\beta}_{1})] ) \nonumber \\
&=\beta_{0} + \bar{X} (\beta_{1} - \beta_{1}) && \text{since $E[\beta_{1}]=\beta_{1}$ and $E[\hat{\beta}_{1}]=\beta_{1}$} \nonumber \\
&= \beta_{0} \nonumber
\end{align}
Therefore, $\hat{\beta}_{0}$ is an unbiased estimator of $\beta_{0}$.

\subsection{Variance of $\hat{\beta}_{1}$ and $\hat{\beta}_{0}$}
Using the assumption that $y_{i}$ are independently distributed, the variance of $\hat{\beta}_{1}$ is
\begin{align}
Var(\hat{\beta}_{1})&= E[ \{\hat{\beta}_{1} - E[\hat{\beta}_{1}] \}^{2}] \\
&= E[ \{\hat{\beta}_{1} - \beta_{1} \}^{2}] && \text{since $E[\hat{\beta}_{1}]=\beta_{1}$} \nonumber
\end{align}

Note equation \eqref{eq06}

\begin{equation}
(\hat{\beta}_{1} - \beta_{1})^{2}=(\sum_{i=1}^{N} k_{i} \epsilon_{i})^{2}
= \sum_{i=1}^{N} k_{i}^{2} \epsilon_{i}^{2}+2\sum_{i=1}^{N-1}\sum_{j=i+1}^{N} k_{i}k_{j}\epsilon_{i}\epsilon_{j}
\end{equation}
Hence,

\begin{align}
E[ \{\hat{\beta}_{1} - \beta_{1} \}^{2}] &= E[\sum_{i=1}^{N} k_{i}^{2} \epsilon_{i}^{2}+2\sum_{i=1}^{N-1}\sum_{j=i+1}^{N} k_{i}k_{j}\epsilon_{i}\epsilon_{j}] \\
&= \sum_{i=1}^{N} k_{i}^{2} E[\epsilon_{i}^{2}|X_{i}] 
+2\sum_{i=1}^{N-1}\sum_{j=i+1}^{N} k_{i}k_{j} E[\epsilon_{i}\epsilon_{j}|X_{i}X_{j}] \nonumber \\
&= \sum_{i=1}^{N} k_{i}^{2} E[\epsilon_{i}^{2}|X_{i}] && \text{since $E[\epsilon_{i}\epsilon_{j}|X_{i}X_{j}]=0$ by assumption} \nonumber \\
&= \sum_{i=1}^{N} k_{i}^{2} \cdot \sigma^{2} = \sigma^{2} \sum_{i=1}^{N} k_{i}^{2} && \text{since $E[\epsilon_{i}^{2}|X_{i}]=\sigma^{2}$ by assumption} \nonumber 
\end{align}

\begin{equation}
\sum_{i=1}^{N} k_{i}^{2} = 
\frac{1}{\{\sum_{i=1}^{N} X_{i} \cdot sinh(\phi_{i})\}^2} \sum_{i=1}^{N} \{ sinh(\phi_{i}) \}^{2}
\end{equation}

\begin{align}
\sum_{i=1}^{N} \{ sinh(\phi_{i}) \}^{2} &= 
\sum_{i=1}^{N} \frac{1}{2} \{ cosh(2 \cdot \phi_{i}) - 1 \} \\
&= \frac{1}{2} \sum_{i=1}^{N} cosh(2 \cdot \phi_{i}) - \frac{N}{2} \nonumber \\
&= \frac{N}{2} \{ \frac{1}{N} \sum_{i=1}^{N} cosh(2 \cdot \phi_{i}) - 1 \} \nonumber
\end{align}

\begin{align}
\sum_{i=1}^{N} \{ sinh(\phi_{i}) \}^{2} = \frac{N}{2}(T-1)
&& \text{where $T=\frac{1}{N} \sum_{i=1}^{N} cosh(2 \cdot \phi_{i})$}
\end{align}

\begin{align}
\sum_{i=1}^{N} X_{i} \cdot sinh(\phi_{i})\ &=
\sum_{i=1}^{N} c \cdot tanh(\theta_{i}) \cdot sinh(\theta_{i}-\theta_{0}) \\ &=
c \sum_{i=1}^{N} tanh(\theta_{i}) \{ sinh(\theta_{i})cosh(\theta_{0}) - cosh(\theta_{i})sinh(\theta_{0}) \} \nonumber \\ &=
c \sum_{i=1}^{N} \{ \frac{sinh(\theta_{i})}{cosh(\theta_{i})} \cdot sinh(\theta_{i})cosh(\theta_{0}) - \frac{sinh(\theta_{i})}{cosh(\theta_{i})} \cdot cosh(\theta_{i})sinh(\theta_{0}) \} \nonumber \\ &=
c \sum_{i=1}^{N} \{ cosh(\theta_{0})\frac{sinh^{2}(\theta_{i})}{cosh(\theta_{i})} - sinh(\theta_{0})sinh(\theta_{i}) \} \nonumber \\ &=
c \sum_{i=1}^{N} [ cosh(\theta_{0}) \{ cosh(\theta_{i}) - \frac{1}{cosh(\theta_{i})} \} - sinh(\theta_{0})sinh(\theta_{i})] \nonumber \\ &=
c [ cosh(\theta_{0}) \{ \sum_{i=1}^{N} cosh(\theta_{i}) - \sum_{i=1}^{N} \frac{1}{cosh(\theta_{i})} \} - sinh(\theta_{0}) \sum_{i=1}^{N} sinh(\theta_{i}) ] \nonumber
\end{align}

Let $C =\frac{1}{N} \sum_{i=1}^{N} cosh(\theta_{i})$, $S=\frac{1}{N} \sum_{i=1}^{N} sinh(\theta_{i})$, and $H=\frac{N}{\sum_{i=1}^{N} \frac {1}{cosh(\theta_{i})}}$.

\begin{align}
tanh(\theta_{0}) &= \frac {S}{C} \\
cosh(\theta_{0}) &= \frac {1}{\sqrt{1-tanh^{2}(\theta_{0})}} = \frac {1}{\sqrt{1-\frac{S^{2}}{C^{2}}}} = \frac {C}{\sqrt{S^{2}-C^{2}}} \\
sinh(\theta_{0}) &= tanh(\theta_{0}) \cdot cosh(\theta_{0})= \frac {S}{\sqrt{S^{2}-C^{2}}}\\
\sum_{i=1}^{N} \frac {1}{cosh(\theta_{i})} &= \frac {N}{H}
\end{align}

\begin{align}
\sum_{i=1}^{N} X_{i} \cdot sinh(\phi_{i})\ &=
c [\frac {C}{\sqrt{S^{2}-C^{2}}} \{N \cdot C -\frac{N}{H} \} - \frac {S}{\sqrt{S^{2}-C^{2}}} \cdot N \cdot S ] \\ &=
\frac {c \cdot N}{\sqrt{S^{2}-C^{2}}}[C^{2}-S^{2}-\frac{C}{H}] \nonumber
\end{align}

Therefore, 
\begin{align}
Var(\hat{\beta}_{1}) &= 
\sigma^{2} \frac {N(T-1)}{2} \frac{S^{2}-C^{2}}{c^{2} \cdot N^{2} (C^{2}-S^{2}-\frac{C}{H})^{2}} 
&= \frac {(S^{2}-C^{2})(T-1)}{2c^{2} \cdot N (C^{2}-S^{2}-\frac{C}{H})^{2}} \sigma^{2} 
\end{align}

The variance of $\hat{\beta}_{0}$ is

\begin{align}
Var(\hat{\beta}_{0})&= Var(\bar{Y} - \hat{\beta}_{1} \cdot \bar{X})
=Var(\bar{Y})+\bar{X}^{2} Var(\hat{\beta}_{1}) \\ &= 
Var(\frac{1}{N} \sum_{i=1}^{N} (\beta_{0}+\beta_{1} \cdot X_{i} + \epsilon_{i}) ) +\bar{X}^{2} Var(\hat{\beta}_{1}) \\ &=
\frac {1}{N^{2}} \cdot N \cdot \sigma^{2} + \bar{X}^{2} Var(\hat{\beta}_{1}) \\ &=
\frac{1}{N} \bigg(1+ \frac{(S^{2}-C^{2})(T-1)\bar{X}^{2}}{2c^{2} (C^{2}-S^{2}-\frac{C}{H})^{2}} \bigg) \sigma^{2}
\end{align}

\subsection{Covariance between $\hat{\beta}_{0}$ and $\hat{\beta}_{1}$}

The covariance between $\hat{\beta}_{0}$ and $\hat{\beta}_{1}$ is

\begin{align}
Cov(\hat{\beta}_{0}, \hat{\beta}_{1}) &= E[(\hat{\beta}_{0}-E[\hat{\beta}_{0}]) (\hat{\beta}_{1} - E[\hat{\beta}_{1}])] \\ &=
E[\{ (\bar{Y}-\hat{\beta}_{1} \bar{X}) - E[\hat{\beta}_{0}] \}(\hat{\beta}_{1} - E[\hat{\beta}_{1}])] && \text{from equation $\eqref{eq07}$} \nonumber \\ &=
E[\{ (\bar{Y}-\hat{\beta}_{1} \bar{X}) - (\bar{Y}-\beta_{1} \bar{X}) \}(\hat{\beta}_{1} - E[\hat{\beta}_{1}])] && \text{since $E[\hat{\beta}_{0}] =\bar{Y}-E[\hat{\beta}_{1}] \bar{X} = \bar{Y}-\beta_{1} \bar{X}$} \nonumber \\ &=
E[\{ (\bar{Y}-\hat{\beta}_{1} \bar{X}) - (\bar{Y}-\beta_{1} \bar{X}) \}(\hat{\beta}_{1} - \beta_{1})] && \text{from equation $\eqref{eq08}$} \nonumber \\ &=
E[-\bar{X} \cdot (\hat{\beta}_{1}-\beta_{1})^{2}] \nonumber \\ &= -\bar{X} \cdot E[ (\hat{\beta}_{1}-\beta_{1})^{2}] \nonumber \\ &= -\bar{X} \cdot Var(\hat{\beta}_{1}) \nonumber
\end{align}

\subsection{Convergence of $\hat{\beta}_{1}$ and $\hat{\beta}_{0}$ to $\hat{\beta}_{1, OLS}$ and $\hat{\beta}_{0, OLS}$ when $c \to \infty$}

Let $X_{0} = c \cdot tanh(\theta_{0})$.

\begin{equation}
X_{0} = c \cdot tanh(\theta_{0}) = c \cdot \frac{\sum_{i=1}^{N} sinh(\theta_{i})}{\sum_{i=1}^{N} cosh(\theta_{i})} = \frac{\sum_{i=1}^{N} c \cdot sinh(\theta_{i})}{\sum_{i=1}^{N} cosh(\theta_{i})}
\end{equation}

\begin{align}
\lim_{c \rightarrow \infty} c \cdot sinh(\theta_{i}) 
&= \lim_{c \rightarrow \infty} c \cdot sinh(tanh^{-1}(\frac{X_{i}}{c}))=X_{i} \\
\lim_{c \rightarrow \infty} cosh(\theta_{i}) 
&= \lim_{c \rightarrow \infty} cosh(tanh^{-1}(\frac{X_{i}}{c})) = 1 \\
\lim_{c \rightarrow \infty} X_{0} &= 
\lim_{c \rightarrow \infty} \frac{\sum_{i=1}^{N} c \cdot sinh(\theta_{i})}{\sum_{i=1}^{N} cosh(\theta_{i})}
=\lim_{c \rightarrow \infty} \frac{\sum_{i=1}^{N} X_{i}}{\sum_{i=1}^{N} 1}
= \frac{1}{N} \sum_{i=1}^{N} X_{i} = \bar{X}
\end{align}

\begin{align}
c \cdot sinh(\phi_{i}) &= c \cdot sinh(\theta_{i} - \theta_{0}) = c \cdot sinh(tanh^{-1}(\frac{X_{i}}{c})-tanh^{-1}(\frac{X_{0}}{c})) \\ &=
c \cdot sinh(tanh^{-1}(\frac{X_{i}}{c})) \cdot cosh(tanh^{-1}(\frac{X_{0}}{c})) \nonumber \\ & \qquad \qquad - c \cdot cosh(tanh^{-1}(\frac{X_{i}}{c})) \cdot sinh(tanh^{-1}(\frac{X_{0}}{c}))
\end{align}

\begin{align}
\lim_{c \rightarrow \infty} c \cdot sinh(\phi_{i}) &= 
\lim_{c \rightarrow \infty} c \cdot sinh(tanh^{-1}(\frac{X_{i}}{c})) \cdot 
\lim_{c \rightarrow \infty} cosh(tanh^{-1}(\frac{X_{0}}{c})) \nonumber \\
& \qquad - \lim_{c \rightarrow \infty} cosh(tanh^{-1}(\frac{X_{i}}{c})) \cdot
\lim_{c \rightarrow \infty} c \cdot sinh(tanh^{-1}(\frac{X_{0}}{c})) \\ 
&= X_{i}-\bar{X}
\end{align}

\begin{align}
\sum_{i=1}^{N} c \cdot X_{i} \cdot sinh(\phi_{i}) &=
\sum_{i=1}^{N} c \cdot \{ (X_{i}-X_{0})+X_{0} \} \cdot sinh(\phi_{i}) \\ &=
\sum_{i=1}^{N} c \cdot (X_{i}-X_{0}) \cdot sinh(\phi_{i}) +
\sum_{i=1}^{N} c \cdot X_{0} \cdot sinh(\phi_{i}) \\ &=
\sum_{i=1}^{N} c \cdot (X_{i}-X_{0}) \cdot sinh(\phi_{i})
\end{align}

\begin{align}
c \cdot (X_{i}-X_{0}) \cdot sinh(\phi_{i}) &=
c \cdot (X_{i}-X_{0}) \cdot sinh(\theta_{i} - \theta_{0}) \\ &=
c \cdot (X_{i}-X_{0}) \cdot sinh(tanh^{-1}(\frac{X_{i}}{c})-tanh^{-1}(\frac{X_{0}}{c})) \\ &=
c \cdot (X_{i}-X_{0}) \cdot sinh(tanh^{-1}(\frac{X_{i}}{c})) \cdot cosh(tanh^{-1}(\frac{X_{0}}{c})) \nonumber \\
& \qquad - c \cdot (X_{i}-X_{0}) \cdot cosh(tanh^{-1}(\frac{X_{i}}{c})) \cdot sinh(tanh^{-1}(\frac{X_{0}}{c}))
\end{align}

\begin{equation}
\begin{split}
\lim_{c \rightarrow \infty} & c \cdot (X_{i}-X_{0}) \cdot sinh(\phi_{i}) \\&=
\lim_{c \rightarrow \infty} (X_{i}-X_{0}) \cdot 
\lim_{c \rightarrow \infty} c \cdot sinh(tanh^{-1}(\frac{X_{i}}{c})) \cdot
\lim_{c \rightarrow \infty} cosh(tanh^{-1}(\frac{X_{0}}{c})) \\ 
& \qquad - \lim_{c \rightarrow \infty} (X_{i}-X_{0}) \cdot 
\lim_{c \rightarrow \infty} cosh(tanh^{-1}(\frac{X_{i}}{c})) \cdot 
\lim_{c \rightarrow \infty} c \cdot sinh(tanh^{-1}(\frac{X_{0}}{c})) \\
&= (X_{i}-\bar{X}) \cdot X_{i} - (X_{i}-\bar{X}) \cdot \bar{X} \\ &= (X_{i}-\bar{X})^{2} 
\end{split}
\end{equation}

\begin{equation}
\begin{split}
\lim_{c \rightarrow \infty} \hat{\beta}_{1} &= 
\lim_{c \rightarrow \infty} \frac{\sum_{i=1}^{N} Y_{i} \cdot sinh(\phi_{i})}{\sum_{i=1}^{N} X_{i} \cdot sinh(\phi_{i})} = 
\lim_{c \rightarrow \infty} \frac{\sum_{i=1}^{N} Y_{i} \cdot c \cdot sinh(\phi_{i})}{\sum_{i=1}^{N} X_{i} \cdot c \cdot sinh(\phi_{i})} \\ &=
\lim_{c \rightarrow \infty} \frac{\sum_{i=1}^{N} Y_{i} \cdot c \cdot sinh(\phi_{i})}{\sum_{i=1}^{N} (X_{i}-X_{0}) \cdot c \cdot sinh(\phi_{i})} \\ &=
\frac{\sum_{i=1}^{N} \lim_{c \rightarrow \infty} Y_{i} \cdot c \cdot sinh(\phi_{i})}{\sum_{i=1}^{N} \lim_{c \rightarrow \infty} (X_{i}-X_{0}) \cdot c \cdot sinh(\phi_{i})}
\\ &=
\frac{\sum_{i=1}^{N} Y_{i} \cdot (X_{i}-\bar{X})}{\sum_{i=1}^{N} (X_{i}-\bar{X})^{2}}
= \hat{\beta}_{1, OLS}
\end{split} 
\end{equation}

Therefore, $\hat{\beta}_{1}$ converges to $\hat{\beta}_{1, OLS}$ when $c \rightarrow \infty$.

\begin{align}
\lim_{c \rightarrow \infty} \hat{\beta}_{0} &= \lim_{c \rightarrow \infty} (\bar{Y} - \hat{\beta}_{1} \cdot \bar{X}) & \text{from equation \eqref{eq07}} \nonumber \\ &=
\bar{Y} - \bar{X} \cdot \lim_{c \rightarrow \infty} \hat{\beta}_{1} = \bar{Y} - \hat{\beta}_{1, OLS} \cdot \bar{X} \nonumber \\ &= \hat{\beta}_{1, OLS}
\end{align}

Therefore, $\hat{\beta}_{0}$ converges to $\hat{\beta}_{0, OLS}$ when $c \rightarrow \infty$.

\subsection{Convergence of $Var(\hat{\beta}_{1})$ and $Var(\hat{\beta}_{0}$) to $Var(\hat{\beta}_{1, OLS})$ and $Var(\hat{\beta}_{0, OLS})$ when $c \to \infty$}

\begin{equation}
\begin{split}
Var(\hat{\beta}_{1}) &= \sigma^{2} \sum_{i=1}^{N} k_{i}^{2} 
= \sigma^{2} \frac{\sum_{i=1}^{N} \{ sinh(\phi_{i}) \}^{2}}{\{\sum_{i=1}^{N} X_{i} \cdot sinh(\phi_{i})\}^2} \\
&= \sigma^{2} \frac{\sum_{i=1}^{N} \{ c \cdot sinh(\phi_{i}) \}^{2}}{\{\sum_{i=1}^{N} X_{i} \cdot c \cdot sinh(\phi_{i})\}^2} 
= \sigma^{2} \frac{\sum_{i=1}^{N} \{ c \cdot sinh(\phi_{i}) \}^{2}}{\{\sum_{i=1}^{N} (X_{i}-X_{0}) \cdot c \cdot sinh(\phi_{i})\}^2} 
\end{split} 
\end{equation}

\begin{equation}
\begin{split}
\lim_{c \rightarrow \infty} Var(\hat{\beta}_{1}) &= 
\sigma^{2} \frac{\sum_{i=1}^{N} \{ \lim_{c \rightarrow \infty} c \cdot sinh(\phi_{i}) \}^{2}}{\{\sum_{i=1}^{N} \lim_{c \rightarrow \infty} (X_{i}-X_{0}) \cdot c \cdot sinh(\phi_{i})\}^2} \\ &=
\sigma^{2} \frac{\sum_{i=1}^{N} (X_{i}-\bar{X})^{2}}
{(\sum_{i=1}^{N} (X_{i}-\bar{X})^{2})^{2}} 
= \frac {\sigma^{2}}{\sum_{i=1}^{N} (X_{i}-\bar{X})^{2}}
\\ &= Var(\hat{\beta}_{1, OLS})
\end{split} 
\end{equation}

Therefore, $Var(\hat{\beta}_{1})$ converges to $Var(\hat{\beta}_{1, OLS})$ when $c \rightarrow \infty$.

\begin{equation}
\begin{split}
\lim_{c \rightarrow \infty} Var(\hat{\beta}_{0})&= \lim_{c \rightarrow \infty} (Var(\bar{Y})+\bar{X}^{2} \cdot Var(\hat{\beta}_{1})) \\ &=
Var(\bar{Y})+\bar{X}^{2} \cdot \lim_{c \rightarrow \infty} Var(\hat{\beta}_{1}) \\ &=
Var(\bar{Y})+\bar{X}^{2} \cdot Var(\hat{\beta}_{1, OLS}) \\ &= Var(\hat{\beta}_{0, OLS})
\end{split}
\end{equation}

Therefore, $Var(\hat{\beta}_{0})$ converges to $Var(\hat{\beta}_{0, OLS})$ when $c \rightarrow \infty$.

\subsection{Convergence of $Cov(\hat{\beta}_{0}, \hat{\beta}_{1})$ to $Cov(\hat{\beta}_{0, OLS}, \hat{\beta}_{1, OLS})$ when $c \to \infty$}

\begin{equation}
\begin{split}
\lim_{c \rightarrow \infty} Cov(\hat{\beta}_{0}, \hat{\beta}_{1}) &= \lim_{c \rightarrow \infty} (-\bar{X} \cdot Var(\hat{\beta}_{1})) = -\bar{X} \cdot \lim_{c \rightarrow \infty} Var(\hat{\beta}_{1})) \\ &=
-\bar{X} \cdot Var(\hat{\beta}_{1, OLS}) \\ &= Cov(\hat{\beta}_{0, OLS}, \hat{\beta}_{1, OLS})
\end{split}
\end{equation}

Therefore, $Cov(\hat{\beta}_{0}, \hat{\beta}_{1})$ converges to $Cov(\hat{\beta}_{0, OLS}, \hat{\beta}_{1, OLS})$ when $c \rightarrow \infty$.

\subsection{Comparison $Var(\hat{\beta}_{1})$ and $Var(\hat{\beta}_{0}$) to $Var(\hat{\beta}_{1, OLS})$ and $Var(\hat{\beta}_{0, OLS})$}

\begin{equation}
\begin{split}
Var(\hat{\beta}_{1}) = \sigma^{2} \frac{\sum_{i=1}^{N} \{ sinh(\phi_{i}) \}^{2}}{\{\sum_{i=1}^{N} X_{i} \cdot sinh(\phi_{i})\}^2} 
\end{split} 
\end{equation}

\begin{equation}
\begin{split}
Var(\hat{\beta}_{1, OLS}) = \sigma^{2} \frac{1}
{\sum_{i=1}^{N} (X_{i}-\bar{X})^{2}} 
\end{split} 
\end{equation}

\begin{equation}
\begin{split}
Var(\hat{\beta}_{1}) - Var(\hat{\beta}_{1, OLS}) &=
\frac {\sigma^{2}}{\{\sum_{i=1}^{N} X_{i} \cdot sinh(\phi_{i})\}^2 \cdot \sum_{i=1}^{N} (X_{i}-\bar{X})^{2}} \\
& \cdot (\sum_{i=1}^{N} \{ sinh(\phi_{i}) \}^{2} \cdot \sum_{i=1}^{N} (X_{i}-\bar{X})^{2}
- \{\sum_{i=1}^{N} X_{i} \cdot sinh(\phi_{i})\}^2 )
\end{split} 
\end{equation}

\begin{align}
\sum_{i=1}^{N} & \{ sinh(\phi_{i}) \}^{2} \cdot \sum_{i=1}^{N} (X_{i}-\bar{X})^{2} 
- \{\sum_{i=1}^{N} X_{i} \cdot sinh(\phi_{i})\}^2 \\ &=
\sum_{i=1}^{N} \{ sinh(\phi_{i}) \}^{2} \cdot \sum_{i=1}^{N} (X_{i}-\bar{X})^{2}
- \{\sum_{i=1}^{N} (X_{i}-\bar{X}) \cdot sinh(\phi_{i})\}^2 \nonumber \\
& > 0 & \text{Cauchy-Schwarz inequality} \nonumber
\end{align}

Therefore, $Var(\hat{\beta}_{1}) > Var(\hat{\beta}_{1, OLS})$.

\begin{equation}
\begin{split}
Var(\hat{\beta}_{0}) &= Var(\bar{Y})+\bar{X}^{2} \cdot Var(\hat{\beta}_{1}) \\ &> Var(\bar{Y})+\bar{X}^{2} \cdot Var(\hat{\beta}_{1, OLS}) = Var(\hat{\beta}_{0, OLS})
\end{split}
\end{equation}

Therefore, $Var(\hat{\beta}_{0}) > Var(\hat{\beta}_{0, OLS})$.

\begin{equation}
\begin{split}
Cov(\hat{\beta}_{0}, \hat{\beta}_{1}) &= -\bar{X} \cdot Var(\hat{\beta}_{1}) \\
&> -\bar{X} \cdot Var(\hat{\beta}_{1, OLS}) = Cov(\hat{\beta}_{0, OLS}, \hat{\beta}_{1, OLS})
\end{split}
\end{equation}

Therefore, $Cov(\hat{\beta}_{0}, \hat{\beta}_{1}) > Cov(\hat{\beta}_{0, OLS}, \hat{\beta}_{1, OLS})$.

\section{Summary and conclusions}
This study investigated the problem posed by using OLS to estimate linear regression parameters when an independent variable is restricted to an open interval, $(-c, c)$, under the context of special relativity. Our investigation revealed that the OLS estimate for the slope parameter is not invariant under the Lorentz velocity transformation.

As an alternative estimator for the parameters of linear regression under special relativity, we proposed an estimator that is invariant under the Lorentz velocity transformation. The proposed estimator was found to be unbiased and converges to the OLS estimator when $c$ approaches to infinity. The variance of the proposed estimator also converges to that of the OLS estimator when $c$ approaches to infinity. Therefore, the proposed estimator can be considered a generalization of the OLS estimator when an independent variable is restricted to an open interval.

The variance of the proposed estimator is larger than that of the OLS estimator, which indicates that there is larger uncertainty when an independent variable is restricted. It shows that hypothesis testing using the OLS estimator and its variance may result in a liberal test when an independent variable is restricted because the confidence interval constructed from the OLS estimator and its variance is narrower than the confidence interval constructed from the proposed estimator and its variance.

There are many circumstances in which independent variables in regression models are restricted to an open interval. Although the proposed estimator may not be applicable to general cases,
our results suggest that one needs to pay attention to the mechanism of how and why the independent variables are restricted and reflect the mechanism in the estimation process. Otherwise, one may obtain misleading estimates, which may result in liberal hypothesis tests.

\newpage
\appendix
\section*{Appendix: Proof of $\sum_{i=1}^{N} X_{i} \cdot sinh(\phi_{i}) > 0$}\label{app}

Let $sinh_{q}(x)$, $cosh_{q}(x)$, and $tanh_{q}(x)$ be q-deformed hyperbolic functions \citep{r6}, as follows.

\begin{equation}
\begin{split}
sinh_{q}(x) &\equiv \frac{e^{x}-q \cdot e^{-x}}{2} \\
cosh_{q}(x) &\equiv \frac{e^{x}+q \cdot e^{-x}}{2} \\
tanh_{q}(x) &\equiv \frac {sinh_{q}(x)}{cosh_{q}(x)} = \frac{e^{x}-q \cdot e^{-x}}{e^{x}+q \cdot e^{-x}}
\end{split}
\end{equation}

The q-deformed hyperbolic functions have the following properties.

\begin{equation}
\begin{split}
cosh^{2}_{q}(x) - sinh^{2}_{q}(x) = q \\
cosh_{q}(x) \geq \sqrt{q} \\
sinh_{q}(x) = 0 \text{ if } x = \frac{1}{2} ln(q)
\end{split}
\end{equation}

\begin{equation}
\begin{split}
sinh(\phi_{i}+\theta_{0}) = \frac{1}{\sqrt{q}} sinh_{q}(\phi_{i}) \\
cosh(\phi_{i}+\theta_{0}) = \frac{1}{\sqrt{q}} cosh_{q}(\phi_{i}) \\
tanh(\phi_{i}+\theta_{0}) = tanh_{q}(\phi_{i}) \\
\end{split}
\end{equation}
where $q = e^{-2 \theta_{0}}$.

\begin{align}
\sum_{i=1}^{N} X_{i} \cdot sinh(\phi_{i}) &=
\sum_{i=1}^{N} c \cdot tanh(\phi_{i}+\theta_{0}) \cdot sinh(\phi_{i}) \nonumber \\ &=
\sum_{i=1}^{N} c \cdot tanh_{q}(\phi_{i}) \cdot sinh(\phi_{i}) \nonumber \\ &=
c \sum_{i=1}^{N} \frac {sinh_{q}(\phi_{i})}{cosh_{q}(\phi_{i})} \cdot sinh(\phi_{i}) \nonumber \\ &=
c \sum_{i=1}^{N} \frac {sinh_{q}(\phi_{i})}{\sqrt{q+sinh^{2}_{q}(\phi_{i})}} \cdot sinh(\phi_{i}) \nonumber \\ 
& > c \sum_{sinh_{q}(\phi_{i}) \neq 0} \frac {sinh_{q}(\phi_{i})}{\sqrt{sinh^{2}_{q}(\phi_{i})}} \cdot sinh(\phi_{i}) & \text{since } q = e^{-2 \theta_{0}} > 0
\end{align}

\begin{equation}
\frac {sinh_{q}(\phi_{i})}{\sqrt{sinh^{2}_{q}(\phi_{i})}} = \begin{cases}
1 & \text{if $\phi_{i} > - \theta_{0}$} , \\
-1 & \text{if $\phi_{i} < - \theta_{0}$}.
\end{cases}
\end{equation}

\begin{equation}
\sum_{sinh_{q}(\phi_{i}) \neq 0} \frac {sinh_{q}(\phi_{i})}{\sqrt{sinh^{2}_{q}(\phi_{i})}} \cdot sinh(\phi_{i}) =  \sum_{\phi_{i} > -\theta_{0}} sinh(\phi_{i}) - \sum_{\phi_{i} < -\theta_{0}} sinh(\phi_{i}) \\
\end{equation}

When $\theta_{0} = 0$ ,
\begin{equation}
\begin{split}
\sum_{\phi_{i} > -\theta_{0}} sinh(\phi_{i}) - \sum_{\phi_{i} < -\theta_{0}} sinh(\phi_{i}) &= 
\sum_{\phi_{i} > 0} sinh(\phi_{i}) - \sum_{\phi_{i} < 0} sinh(\phi_{i}) \\
&= 2\sum_{\phi_{i} > 0} sinh(\phi_{i}) > 0
\end{split}
\end{equation}

When $\theta_{0} > 0$ ,
\begin{equation}
\begin{split}
\sum_{\phi_{i} > -\theta_{0}} sinh(\phi_{i}) &- \sum_{\phi_{i} < -\theta_{0}} sinh(\phi_{i}) \\ &= 
\sum_{\phi_{i} > 0} sinh(\phi_{i}) + \sum_{- \theta_{0} < \phi_{i} < 0 } sinh(\phi_{i}) - \sum_{ \phi_{i} \leq - \theta_{0}} sinh(\phi_{i}) \\ &=
- \sum_{\phi_{i} \leq - \theta_{0}} sinh(\phi_{i}) + \{ \sum_{- \theta_{0} < \phi_{i} < 0} sinh(\phi_{i}) - \sum_{\phi_{i} < 0} sinh(\phi_{i}) \} \\ &=
-2 \sum_{\phi_{i} \leq - \theta_{0}} sinh(\phi_{i}) \geq 0
\end{split}
\end{equation}

When $\theta_{0} < 0$ ,
\begin{equation}
\begin{split}
\sum_{\phi_{i} > -\theta_{0}} sinh(\phi_{i}) &- \sum_{\phi_{i} < -\theta_{0}} sinh(\phi_{i}) \\ &= 
\sum_{\phi_{i} > - \theta_{0}} sinh(\phi_{i}) - \sum_{0 < \phi_{i} < - \theta_{0}} sinh(\phi_{i}) - \sum_{ \phi_{i} < 0} sinh(\phi_{i}) \\ &=
\sum_{\phi_{i} > - \theta_{0}} sinh(\phi_{i}) + \{ \sum_{\phi_{i} > 0} sinh(\phi_{i}) - \sum_{0 < \phi_{i} < - \theta_{0}} sinh(\phi_{i}) \} \\ &=
2 \sum_{\phi_{i} > - \theta_{0}} sinh(\phi_{i}) \geq 0
\end{split}
\end{equation}

Therefore, $\sum_{i=1}^{N} X_{i} \cdot sinh(\phi_{i}) > 0$ for all $\theta_{0}$.



%
%
%
%
\newpage
\bibliographystyle{Chicago}

\bibliography{Bibliography-LRSR}
\end{document}